\documentclass[12pt]{iopart}
\usepackage{graphicx}

\begin{document}

\title[Entanglement and discord in open quantum systems]{Entanglement and discord in two-mode Gaussian open quantum systems}

\author{Aurelian Isar}

\address{National Institute of Physics and Nuclear Engineering,
P.O.Box MG-6, Bucharest-Magurele, Romania}
\ead{isar@theory.nipne.ro}
\begin{abstract}

In the framework of the theory of open systems, we give a description of quantum entanglement and quantum discord for two non-interacting modes embedded in a thermal environment. We describe the evolution of entanglement in terms of the covariance matrix for Gaussian input states. For all values of the temperature of the thermal reservoir, an initial separable squeezed thermal state remains separable for all times. In the case of an entangled initial squeezed thermal state, entanglement suppression (entanglement sudden death) takes place, for non-zero temperatures of the environment. The Gaussian quantum discord, which is a measure of all quantum correlations in the state, including entanglement, decays asymptotically in time.
\end{abstract}


\section{Introduction}

In the framework of the theory of open systems based on completely positive quantum dynamical semigroups, we study the dynamics of quantum entanglement and quantum discord of a subsystem consisting of two uncoupled modes interacting with a common thermal bath. The initial state of the subsystem is taken of Gaussian form and the evolution under the dynamical semigroup assures the preservation in time of the Gaussian form of the state. In Sec. 2 we give the solution of the evolution equation for the covariance matrix corresponding to a generic two-mode Gaussian state of the two uncoupled modes interacting with the environment. We investigate in Sec. 3 the dynamics of entanglement, by analyzing the time evolution of the logarithmic negativity, which characterizes the degree of entanglement of the state. We study also the time evolution of the Gaussian quantum discord, which is a measure of all quantum correlations in the bipartite state including entanglement. Summary is given in Sec. 4.

\section{Equations of motion for two modes interacting with an environment}

We study the dynamics
of a subsystem composed of two non-interacting modes in weak interaction with a thermal environment. In the axiomatic formalism
based on completely positive quantum dynamical semigroups, the Markovian irreversible time
evolution of an open system is described by the Lindblad-Kossakowski master equation.
We are interested in the set of Gaussian states, therefore we introduce such quantum
dynamical semigroups that preserve this set during time evolution of the system.

The two-mode Gaussian state is entirely specified by its
covariance matrix, which is a real,
symmetric and positive matrix:
\begin{eqnarray}\sigma(t)=\left(\matrix{\sigma_{xx}(t)&\sigma_{xp_x}(t) &\sigma_{xy}(t)&
\sigma_{xp_y}(t)\cr \sigma_{xp_x}(t)&\sigma_{p_xp_x}(t)&\sigma_{yp_x}(t)
&\sigma_{p_xp_y}(t)\cr \sigma_{xy}(t)&\sigma_{yp_x}(t)&\sigma_{yy}(t)
&\sigma_{yp_y}(t)\cr \sigma_{xp_y}(t)&\sigma_{p_xp_y}(t)&\sigma_{yp_y}(t)
&\sigma_{p_yp_y}(t)}\right)\equiv\left(\begin{array}{cc}A&C\\
C^{\rm T}&B \end{array}\right),\label{covar} \end{eqnarray}
where $A$, $B$ and $C$ are $2\times 2$ Hermitian matrices. $A$
and $B$ denote the symmetric covariance matrices for the
individual reduced one-mode states, while the matrix $C$
contains the cross-correlations between modes.
The equations of motion for the quantum correlations of the coordinates
$x,y$ and momenta $p_x,p_y$ of the two modes are the following ($\rm T$ denotes the transposed matrix) \cite{san}:
\begin{eqnarray}{d \sigma(t)\over
dt} = Y \sigma(t) + \sigma(t) Y^{\rm T}+2 D,\label{vareq}\end{eqnarray} where
\begin{eqnarray}
Y=\left(\matrix{ -\lambda&1/m&0 &0\cr -m\omega_1^2&-\lambda&0&
0\cr 0&0&-\lambda&1/m \cr 0&0&-m\omega_2^2&-\lambda}\right),\\
D=\left(\matrix{
D_{xx}& D_{xp_x} &D_{xy}& D_{xp_y} \cr D_{xp_x}&D_{p_x p_x}&
D_{yp_x}&D_{p_x p_y} \cr D_{xy}& D_{y p_x}&D_{yy}& D_{y p_y}
\cr D_{xp_y} &D_{p_x p_y}& D_{yp_y} &D_{p_y p_y}} \right),\end{eqnarray}
$m$ is the mass, $\omega_1$ and $\omega_2$
are the frequencies of the non-resonant modes, and diffusion coefficients $D_{xx}, D_{xp_x},$... and dissipation constant $\lambda$ are real quantities.

The time-dependent
solution of Eq. (\ref{vareq}) is given by \cite{san}
\begin{eqnarray}\sigma(t)= M(t)[\sigma(0)-\sigma(\infty)] M^{\rm
T}(t)+\sigma(\infty),\label{covart}\end{eqnarray} where the matrix $M(t)=\exp(Yt)$ has to fulfill
the condition $\lim_{t\to\infty} M(t) = 0.$
The values at infinity are obtained
from the equation \begin{eqnarray}
Y\sigma(\infty)+\sigma(\infty) Y^{\rm T}=-2 D.\label{covarinf}\end{eqnarray}

\section{Dynamics of continuous variable entanglement and discord}

\subsection{Time evolution of entanglement and logarithmic negativity}

In order to quantify the degree of entanglement of the two-mode states it is suitable to use the logarithmic negativity. For a Gaussian density operator, the logarithmic negativity is given by
$
E_N={\rm max}\{0,-\log_2 2\tilde\nu_-\},
$
where $\tilde\nu_-$ is the smallest of the two symplectic eigenvalues of the partial transpose $\tilde{{\sigma}}$ of the 2-mode covariance matrix $\sigma$.
In our model, the logarithmic negativity is calculated as \cite{aeur,aijqi,ascri,aosid}
\begin{eqnarray}E_N(t)={\rm max}\{0,-\frac{1}{2}\log_2[4g(\sigma(t))]\}, \end{eqnarray} where \begin{eqnarray}g(\sigma(t))=\frac{1}{2}(\det A +\det
B)-\det C\nonumber\\
-\left({\left[\frac{1}{2}(\det A+\det B)-\det
C\right]^2-\det\sigma(t)}\right)^{1/2}.\end{eqnarray}

We assume that the initial Gaussian state is the two-mode squeezed thermal state, with the covariance matrix of the form \cite{mar}
\begin{eqnarray}\sigma_{st}(0)=\frac{1}{2}\left(\matrix{a&0&c&0\cr
0&a&0&-c\cr
c&0&b&0\cr
0&-c&0&b}\right),\label{ini1} \end{eqnarray}
with the matrix elements given by
\begin{eqnarray}a=n_1 \cosh^2 r + n_2 \sinh^2 r + \frac{1}{2} \cosh 2r,\\
b=n_1 \sinh^2 r + n_2 \cosh^2 r + \frac{1}{2} \cosh 2r,\\
c=\frac{1}{2}(n_1 + n_2 + 1) \sinh 2r,\label{ini2}
\end{eqnarray}
where $n_1,n_2$ are the average numbers of thermal photons associated with the two modes and $r$ denotes the squeezing parameter.
In the particular case $n_1=0$ and $n_2=0$, (\ref{ini1}) becomes the covariance matrix of the two-mode squeezed vacuum state. A two-mode squeezed thermal state is entangled when the squeezing parameter $r$ satisfies the inequality $r>r_s$ \cite{mar},
where \begin{eqnarray} \cosh^2 r_s=\frac{(n_1+1)(n_2+1)}{ n_1+n_2+1}. \end{eqnarray}

We suppose that the asymptotic state of the considered open system is a Gibbs state corresponding to two independent modes in thermal equilibrium at temperature $T.$ Then the quantum diffusion coefficients have the following form (we
put $\hbar=1$) \cite{rev}:
\begin{eqnarray}m\omega_1 D_{xx}=\frac{D_{p_xp_x}}{m\omega_1}=\frac{\lambda}{2}\coth\frac{\omega_1}{2kT},~
m\omega_2 D_{yy}=\frac{D_{p_yp_y}}{m\omega_2}=\frac{\lambda}{2}\coth\frac{\omega_2}{2kT},\label{envcoe}\\
D_{xp_x}=D_{yp_y}=D_{xy}=D_{p_xp_y}=D_{xp_y}=D_{yp_x}=0.\nonumber\end{eqnarray}

The calculations with separable initial squeezed thermal states show that these states remain separable for all values of the temperature $T$ and for all times.
The evolution of entangled initial squeezed thermal states with the covariance matrix given by Eq. (\ref{ini1}) is illustrated in Figure 1, where we represent the dependence of the logarithmic negativity $E_N(t)$ on time $t$ and temperature $T$ for the case of an initial symmetric Gaussian state, when $a=b$.
We observe that for a non-zero temperature $T,$ at certain finite moment of time, which depends on $T,$ $E_N(t)$ becomes zero and therefore the state becomes separable. This is the so-called phenomenon of entanglement sudden death. It is in contrast to the quantum decoherence, during which the loss of quantum coherence is usually gradual \cite{aphysa,arus}. For $T=0,$ $E_N(t)$ remains strictly positive for finite times and tends asymptotically to 0 for $t\to \infty.$ Therefore, only for zero temperature of the thermal bath the initial entangled state remains entangled for finite times and this state tends asymptotically to a separable one for infinitely large time. Dissipation favorizes the phenomenon of entanglement sudden death -- with increasing the dissipation parameter $\lambda,$ the entanglement suppression happens earlier. The same qualitative behaviour of the time evolution of entanglement was obtained previously \cite{ascri2,aosid1} in the particular case $n_1=0$ and $n_2=0$ corresponding to an initial two-mode squeezed vacuum state.

\begin{figure}
\resizebox{0.5\columnwidth}{!}
{
\includegraphics{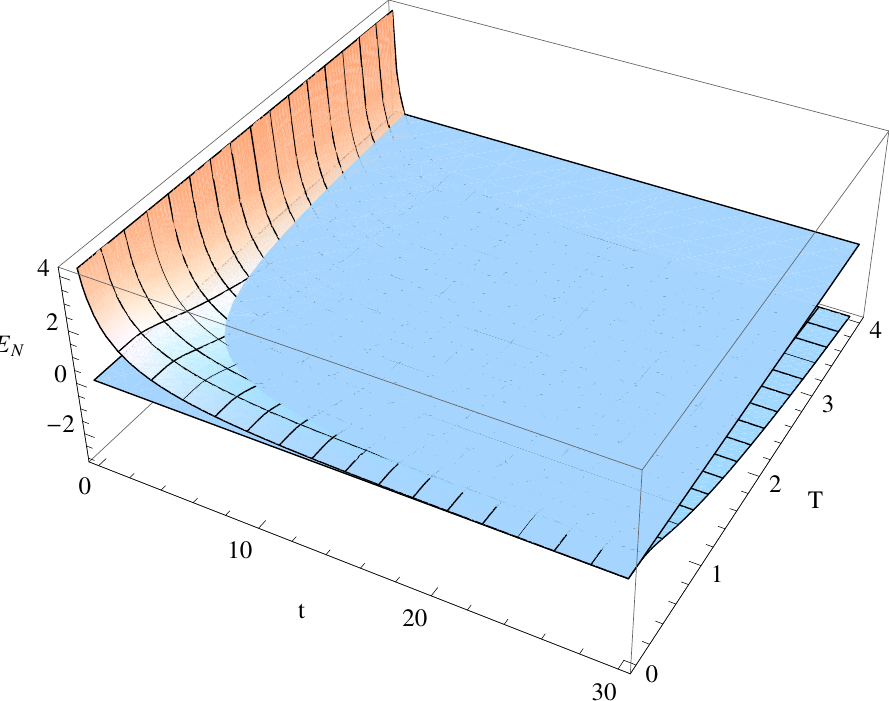}
}
\caption{Logarithmic negativity $E_N$ versus time $t$ and temperature $T$ for an entangled initial symmetric squeezed thermal state with squeezing parameter $r=2$, $n_1=1,n_2=1$ and $\lambda=0.1, \omega_1=\omega_2=1.$ We take $m=\hbar=k=1.$
}
\label{fig:1}
\end{figure}

\subsection{Gaussian quantum discord}

Quantum discord was introduced \cite{zur} as a measure of all quantum
correlations in a bipartite state, including -- but not restricted to -- entanglement.
Originally the quantum discord was defined and evaluated mainly for finite dimensional systems. Very recently \cite{par,ade} the notion of discord has been extended to the domain of
continuous variable systems, in particular to the analysis of
bipartite systems described by two-mode Gaussian states.
Closed formulas have been derived for bipartite thermal squeezed states \cite{par} and for all two-mode Gaussian states \cite{ade}.
The Gaussian quantum discord of a general two-mode Gaussian state
can be defined as the quantum discord where the conditional entropy is restricted to generalized Gaussian
positive operator valued measurements (POVM) on the mode 2, and in terms of symplectic invariants it is given by (the symmetry between the two modes 1 and 2 is broken) \cite{ade}
\begin{eqnarray}
D=f(\sqrt{\beta})-f(\nu_-) - f(\nu_+) + f(\sqrt{\varepsilon}),
\label{disc}
\end{eqnarray}
where \begin{eqnarray}f(x) =\frac{x+1}{2} \log\frac{x+1}{2} -\frac{x-1}{2} \log\frac{x-1}{2},\end{eqnarray}
\begin{eqnarray}\label{infdet}
\varepsilon=
 &  &
\hspace*{-.1cm}
\left\{  \hspace*{-.5cm}  \begin{array}{rcl}& &\begin{array}{c}\displaystyle{\frac{{2 \gamma^2+(\beta-1)(\delta-\alpha)
+2 |\gamma| \sqrt{\gamma^2+(\beta-1) (\delta-\alpha)}}}{{(\beta-1){}^2}}}\end{array},\\& &\qquad
\hbox{if}~~(\delta-\alpha\beta)^2 \le (\beta+1)\gamma^2 (\alpha +\delta)\\ \\& &
\begin{array}{c}\displaystyle{\frac{{\alpha\beta-\gamma^2+\delta-\sqrt{\gamma^4+(\delta-\alpha\beta){}^2-
2\gamma^2(\delta+\alpha\beta)}}}{{2\beta}}}\end{array}, \\& & \qquad
\hbox{otherwise,} \end{array} \right.
\end{eqnarray}
\begin{eqnarray}\alpha=4\det A,~~~\beta=4\det B,~~~\gamma=4\det C,~~~\delta=16\det\sigma,\end{eqnarray}
and $\nu_\mp$ are the symplectic eigenvalues of the state, given by
\begin{eqnarray}2{\nu}_{\mp}^2 ={\Delta}\mp\sqrt{{\Delta}^2
-4\det\sigma},
\end{eqnarray}
where
$\Delta=\det A+\det B+2\det C.$
Notice that Gaussian quantum discord only depends on $|\det C|$, i.e., entangled ($\det C<0$) and separable states are treated on equal footing.

\begin{figure}
\resizebox{0.99\columnwidth}{!}
{
\includegraphics{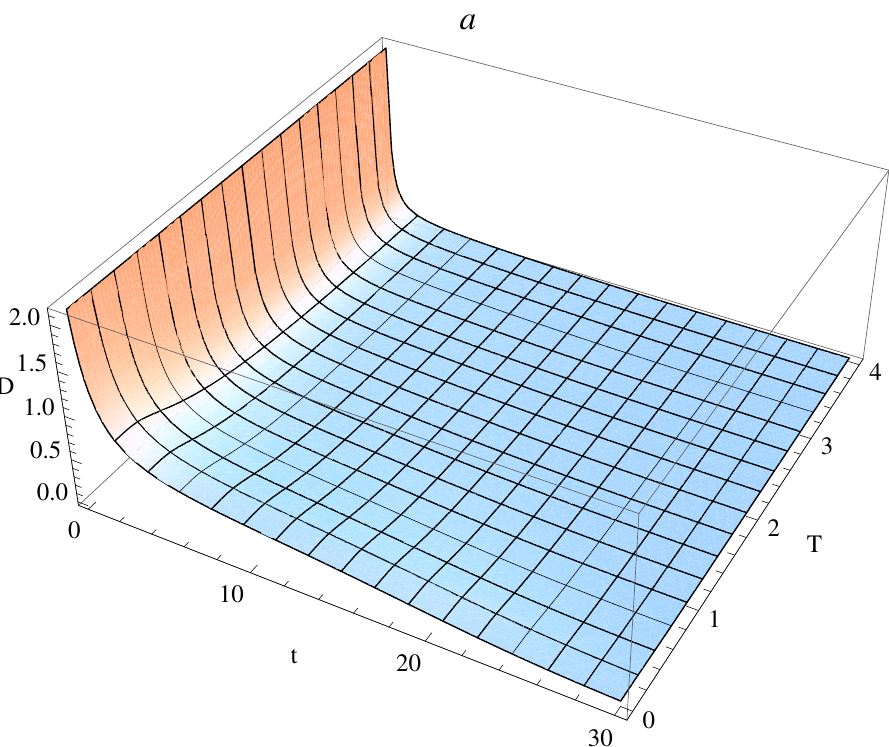}
\includegraphics{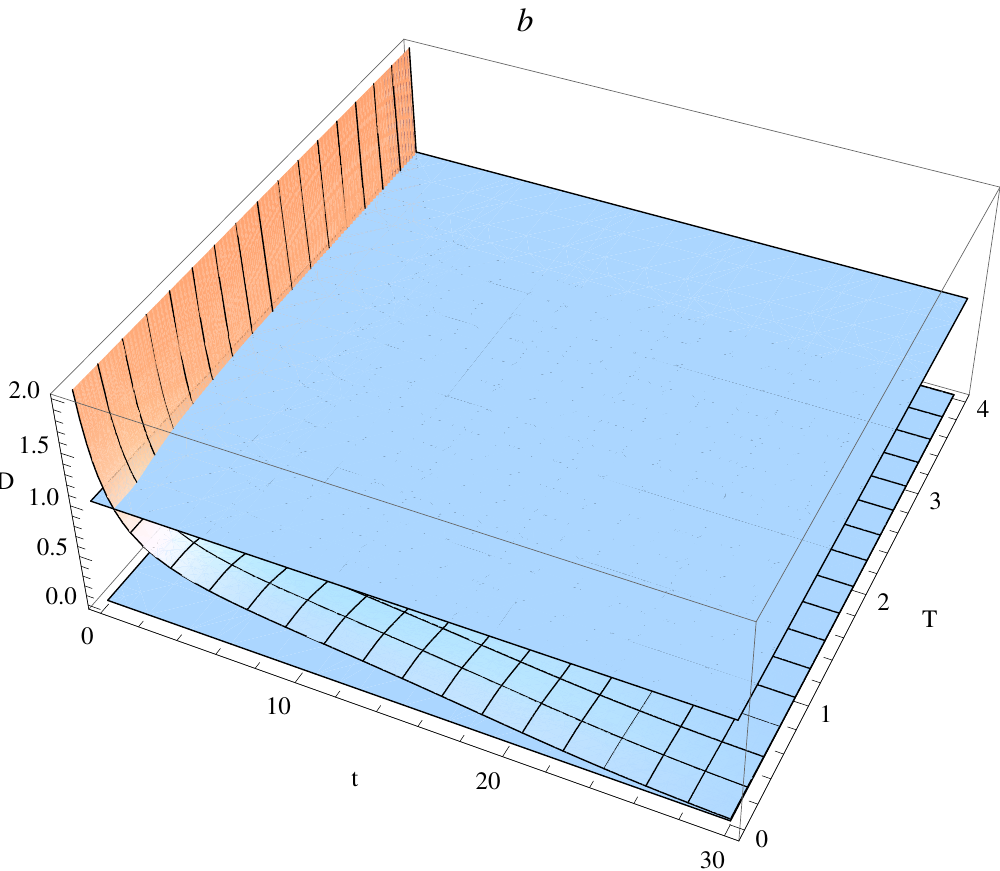}
}
\caption{(a) Gaussian quantum discord $D$ versus time $t$ and temperature $T$ for an entangled initial symmetric squeezed thermal state with squeezing parameter $r=2$, $n_1=n_2=1$ and $\lambda=0.1, \omega_1=\omega_2=1.$ We take $m=\hbar=k=1.$ (b) Same as in (a), with the explicit thresholds shown for $D=0$ and $D=1.$
}
\label{fig:2}
\end{figure}

The evolution of the Gaussian quantum discord $D$ is illustrated in Figure 2, where we represent the dependence of $D$ on time $t$ and temperature $T$ for an entangled initial symmetric Gaussian state, taken of the form of a two-mode squeezed thermal state (\ref{ini1}), for such values of the parameters which satisfy for all times the first condition in formula (\ref{infdet}). For entangled initial states the Gaussian discord remains strictly positive in time and in the limit of infinite time it tends asymptotically to zero, corresponding to the thermal product (separable) state, with no correlations at all. The Gaussian discord has nonzero values for all finite times and this fact certifies the existence of nonclassical correlations in two-mode Gaussian states, either separable or entangled.

From Figure 2 we notice that, in concordance with the general properties of the Gaussian quantum discord \cite{ade}, the states can be either separable or entangled for $D\le 1$ and all the states above the threshold $D=1$ are entangled.
We also notice that the decay of quantum discord is stronger when the temperature $T$ is increasing.

\section{Summary}

In the framework of the theory of open systems we investigated the Markovian dynamics of quantum correlations for a subsystem
composed of two non-interacting modes embedded in a thermal bath.
We have described the time evolution of the logarithmic negativity in terms
of the covariance matrix for squeezed thermal states, for the case when the asymptotic state is a Gibbs state corresponding to two independent quantum harmonic oscillators in thermal equilibrium. For all values of the temperature of the thermal reservoir, an initial separable Gaussian state remains separable for all times. In the case of an entangled initial squeezed thermal state, entanglement sudden death takes place for non-zero temperatures of the environment. Only for a zero temperature of the thermal bath the initial entangled state remains entangled for finite times, but in the limit of infinite time it evolves asymptotically to an equilibrium state which is always separable.
The Gaussian quantum discord decreases asymptotically in time. This is contrast with the sudden death of entanglement. The time evolution
of quantum discord is very similar to that of the entanglement
before the sudden suppression of the entanglement. After the sudden death of the entanglement, the nonzero values of quantum discord
quantify the nonclassical correlations for separable mixed states.

\ack

The author acknowledges the financial support received from the Romanian Ministry of Education and Research, through
the Projects IDEI 497/2009 and PN 09 37 01 02/2010.


\section*{References}

\begin{thebibliography}{99}

\bibitem{san} Sandulescu A, Scutaru H and Scheid W 1987 {\it J. Phys. A: Math. Gen.} \textbf{20} 2121

\bibitem{aeur} Isar A 2008 {\it Eur. J. Phys. Special Topics} \textbf{160} 225

\bibitem{aijqi} Isar A 2008 {\it Int. J. Quantum Inf.} \textbf{6} 689

\bibitem{ascri} Isar A 2009 {\it Phys. Scr., Topical Issue} \textbf{135} 014033

\bibitem{aosid} Isar A 2009 {\it Open Sys. Inf. Dynamics} \textbf{16} 205

\bibitem{mar} Marian P, Marian TA and Scutaru H 2003 {\it Phys. Rev. A} \textbf{68} 062309

\bibitem{rev} Isar A, Sandulescu A, Scutaru H, Stefanescu E and Scheid W 1994 {\it Int. J. Mod. Phys.} E \textbf{3} 635

\bibitem{aphysa} Isar A and Scheid W 2007 {\it Physica} A \textbf{373} 298

\bibitem{arus} Isar A 2007 {\it J. Russ. Laser Res.} \textbf{28} 439

\bibitem{ascri2} Isar A 2011 {\it Phys. Scr.} T\textbf{143} 014012

\bibitem{aosid1} Isar A 2011 {\it Open Sys. Inf. Dynamics} \textbf{18} 1

\bibitem{zur} Zurek WH 2000 {\it Annalen der Physik} (Leipzig) \textbf{9} 853

\bibitem{par} Giorda P and Paris MGA 2010 {\it Phys. Rev. Lett.} \textbf{105} 020503

\bibitem{ade} Adesso G and Datta A 2010 {\it Phys. Rev. Lett.} \textbf{105}  030501

\end{thebibliography}
\end{document}